\documentstyle[aps,prl,twocolumn,epsf]{revtex}

\begin{document}
%\hoffet-0.5cm
\tighten
\newcommand{\be}{\begin{equation}}
\newcommand{\ee}{\end{equation}}
\newcommand{\bea}{\begin{eqnarray}}
\newcommand{\eea}{\end{eqnarray}}
\newcommand{\r}{\rangle }
\newcommand{\la}{\langle }
\title{ A Diagrammatic Interpretation of the Boltzmann Equation}
%\title{Nolinear Response Dynamics  from Classical Transport Theory
%and Quantum Field Theory}
\author{ M.E. Carrington${}^{a,b}$, Hou Defu${}^{a,b,c}$ and R.
Kobes${}^{b,d}$}

 \address{ ${}^a$ Department of Physics, Brandon University, Brandon, 
Manitoba,
R7A 6A9 Canada\\
 ${}^b$  Winnipeg Institute for Theoretical Physics, Winnipeg, Manitoba \\
${}^c$ Institute of Particle Physics, Huazhong Normal University, 430070 Wuhan,
China \\
${}^d$ University of Winnipeg, Winnipeg, Manitoba, R3B 2E9 Canada }

\maketitle

\begin{abstract}

We study nonlinear response in weakly coupled nonequilibrium $\phi^4$ theory
in the context of both classical transport theory and real time 
quantum field theory, based
on a generalized Kubo formula which we derive. A novel connection 
 between these two approaches is established which provides a diagrammatic 
interpretation of the Boltzmann equation.


\end{abstract}

\narrowtext

%%%%%%%%%%%%%%%%%%%%%%%%%%%%%%%%%%%%%%%%%%%%%%%%%%%%%%%%%%%%%%%%%%

%\sect

    Fluctuations occur in a system perturbed slightly away from equilibrium.
The responses to these fluctuations are described by transport coefficients
 \cite{Groot}.
The investigation of transport coefficients in high temperature
gauge theories is important in cosmological applications
such as electroweak baryogenesis and in the context of heavy ion
collisions \cite{hvi}.
 There are two basic methods to calculate transport coefficients: transport
theory and linear response theory
\cite{MartinK,kubo,Hosoya}.  Using the transport
theory method one starts from a local equilibrium form for the distribution
function and performs an expansion in the gradient of the four-velocity field.
The coefficients of this expansion are determined from the 
 classical Boltzmann equation \cite{jeon1}.
In the response theory approach one divides the Hamiltonian into a bare piece
and a perturbative piece that is linear in the gradient of the four-velocity
field.  One uses standard perturbation theory to obtain the Kubo formula for
the viscosity in terms of retarded Green functions \cite{kubo,Hosoya},
which are then evaluated using equilibrium quantum field theory.  As is
typical in finite temperature field theory, it is not sufficient to calculate
perturbatively in the coupling constant: there are certain infinite sets of
diagrams that contribute at the same order in perturbation theory and have to
be resummed \cite{pisarski,LAD}.

To date, most calculations of transport coefficients have 
been done to the order of linear 
response. In many physical situations however nonlinear response can be 
important \cite{bdc,jackiw,blz,ASY}. In this Letter, we study nonlinear response 
using transport theory
and using quantum field theory, and explain the connection between these 
approaches. We perform a Chapman-Enskog expansion of the Boltzmann equation 
keeping up to quadratic contributions.  We obtain a generalized nonlinear 
Kubo formula, and a set of integral equations
which resum ladder and extended ladder diagrams. We show that these 
two equations have exactly the same structure, and thus provide a diagrammatic 
interpretation of the Chapman-Enskog expansion of the Boltzmann equation, up 
to quadratic order.
% We leave for a longer publication a more extensive discussion of the details \cite{mhrkb}.

We start from the definition of shear viscosity. In a system that is out of 
equilibrium, the existence of gradients
in thermodynamic parameters like the temperature and the four dimensional
velocity field give rise to thermodynamic forces which
lead to  deviations from the equilibrium expectation value of the
viscous shear stress:
\bea
&& \delta \langle  \pi_{\mu\nu}\rangle =\eta^{(1)} H_{\mu\nu} + \eta^{(2)}
H^{T2}_{\mu\nu} + \cdots  \label{DEF} \\
&& H_{\mu\nu} = \partial_\mu u_\nu + \partial_\nu u_\nu - \frac{2}{3}
\Delta_{\mu\nu} \Delta_{\rho\sigma}
\partial^\rho  u^\sigma \nonumber \\
&&
H_{\mu\nu}^{T2}:=H_{\mu\rho}H^\rho_{~\nu}-
\frac{1}{3}\Delta_{\mu\nu}H_{\rho\sigma}
H^{\rho\sigma} \nonumber
\eea
where
$u_\mu(x)$ is the four dimensional
four-velocity field which satisfies $u^\mu(x) u_\mu(x)=1$.
$\eta^{(1)}$ and $\eta^{(2)}$ are the coefficients of the terms that are linear
and quadratic respectively in the gradient of the four-velocity. The first
coefficient  is the usual shear viscosity.  The second has has not been 
widely discussed in the literature -- we will call it the quadratic shear 
viscous response.

The  Boltzmann equation can be used to calculate transport
properties for weak coupling $\lambda \phi^4$ theory with zero chemical potential\cite{jeon1}.
 We introduce a phase space distribution function
$f(x,\underline{k})$ (the underlined momenta  are on
shell).  The form of
$f(x,\underline{k})$ in local equilibrium is,
\be
f^{(0)}=\frac{1}{e^{\beta(x) u_\mu(x) 
\underline{k}^\mu}-1}:=n_k\,;~~N_k:=1+2n_k
\,.\label{fequib}
\ee
We expand $ f$  around $f^0$ using a
gradient expansion in the local rest frame where
$\vec{u}(x)=0$.   We keep only linear terms that contain one
power of $H_{\mu\nu}$ and quadratic terms that contain two powers of 
$H_{\mu\nu}$,
since these are the only terms that contribute to the viscosity coefficients we
are trying to calculate. We write,
\bea
f=f^{(0)} + f^{(1)} + f^{(2)} + \cdots
\label{expf}
\eea
with,
\bea
f^{(1)} \sim \underline{k}_\mu \partial^\mu f^{(0)}\,;~~~~f^{(2)} \sim
\underline{k}_\mu \partial^\mu f^{(1)}.
\eea
Using (\ref{fequib}) we obtain,
\bea
 && f^{(1)} := - n_k(1+n_k) \phi_k\,;~~ \phi_k =\beta
\frac{1}{2}B_{ij}(\underline{k})H_{ij}  \label{def2} \\
&&f^{(2)} := n_k(1+n_k)N_k \theta_k\,;~~\theta_k:= \beta^2 \frac{1}{4}
C_{ijlm}(\underline{k}) H_{ij}H_{lm}\,. \nonumber
\eea
We write
\bea
&& B(\underline{k})_{ij} = \hat I_{lm}(k)
B(\underline{k})\,;~~C_{ijlm}(\underline{k}) =  \hat I_{ij}(k) \hat I_{lm}(k)
C(\underline{k}) \nonumber \\
&& \hat I_{ij}(k) =
(\hat k_i \hat k_j-\frac{1}{3}\delta_{ij})\,;~~I_{ij}(k) = 
k^2\hat I_{ij}(k)\,. \label{def4}
\eea

The viscous shear stress  tensor is given by
\bea
\langle  \pi_{ij}\rangle  = \int \frac{d^3 k}{(2\pi)^3 2\omega_k}  f\, (k_i
k_j-\frac{1}{3}\delta_{ij} k^2)\,.
\eea
Using the expansion (\ref{expf}),  (\ref{def2}) and (\ref{def4}) and rotational
invariance we obtain \cite{mhrkb},
\bea
&& \delta\langle \pi_{ij}\rangle =  -\frac{\beta}{15} \int \frac{d^3
k}{(2\pi)^3 2\omega_k} n_k(1+n_k) k^2B(\underline{k})\,H_{ij}
\nonumber
\\
&&~~~~
+\frac{2\beta^2}{105}  \int \frac{d^3 k}{(2\pi)^3 2\omega_k} 
[n_k(1+n_k)N_k] k^2
 C(\underline{k}) H^{T2}_{ij}\,.  \nonumber
\eea
Comparing with (\ref{DEF}) we have,
\bea
&&\eta^{(1)} = \frac{\beta}{15} \int \frac{d^3 k}{(2\pi)^3 2\omega_k}
n_k(1+n_k) k^2B(\underline{k}) \label{ttf11} \\
&& \eta^{(2)} = \frac{2\beta^2}{105}  \int \frac{d^3 k}{(2\pi)^3 2\omega_k}
[n_k(1+n_k)N_k] k^2  C(\underline{k})\,.\label{ttf}
\eea

  Next  we will show that $B(\underline{k})$ and
$C(\underline{k})$ can be
obtained from the first two equations in the hierarchy of equations obtained
from the gradient expansion of the Boltzmann equation, which has the form:
\bea
\underline{k}_\mu \partial^\mu f(x,\underline{k}) = {\cal C}[f] \,.
\label{BT}
\eea
The collision term is ${\cal C}[f]: = \frac{1}{2} \int_{123} d \,
\Gamma_{12\leftrightarrow 3k}[f_1 f_2 (1+f_3)(1+f_k) - 
(1+f_1)(1+f_2) f_3 f_k ]$ 
with $f_i:= f(x,\underline{p}_i)$, $f_k:=f(x,\underline{k})$.  The symbol 
$d \, \Gamma_{12\leftrightarrow 3k}$ 
represents the differential transition rate for
particles of
momentum $P_1$ and $P_2$ to scatter into momenta $P_3$ and $K$.

The first order equation is:
\be
\underline{k}^\mu\partial_\mu f^0 (x,\underline{k})={\cal C}[f^{(0)};f^{(1)}]
\label{foe}
\ee
where we keep terms linear in $f^{(1)}$ on the right hand side.
Using  (\ref{def2}) and comparing the
coefficients of $H_{ij}$ on both sides of (\ref{foe}) 
we obtain \cite{jeon1,meg},
\bea
&&I_{ij}(k)=\frac{1}{2} \int_{123} d \,\Gamma_{12\leftrightarrow 3k}
d_n [B_{ij}(\underline{p}_1) 
\nonumber
\\
&&~~~~~
+ B_{ij}(\underline{p}_2) - B_{ij}(\underline{k}) - B_{ij}(\underline{p}_3)\,]
\label{Bint}
\eea
where $d_n=(1+n_1)(1+n_2)n_3/(1+ n_k)$.
%This inhomogeneous linear integral equation can be solved
%self-consistently to obtain the function $B_{ij}(\underline{k})$.

%\subsubsection{Second Order Boltzmann Equation}
The second order contribution to (\ref{BT}) is,
\be
\underline{k}^\mu\partial_\mu f^{(1)} (x,\underline{k})={\cal
C}[f^{(0)};f^{(1)}; f^{(2)}]
\label{b2}
\ee
where we keep terms linear in $f^{(2)}$ and quadratic in $f^{(1)}$ on the right
hand side.
Using (\ref{def2}) and comparing the
coefficients of $H_{ij}H_{lm}$ on both sides  we obtain,
\bea
   N_k&&I_{ij}(k) B_{lm}(\underline{k}) = \frac{1}{2} \int_{123} d
\,\Gamma_{12\leftrightarrow 3k}d_n\,\{ [ N_1 C_{ijlm}(\underline{p}_1) 
\nonumber\\
&& +
N_2 C_{ijlm}(\underline{p}_2) 
- N_k C_{ijlm} (\underline{k}) - N_3
C_{ijlm}(\underline{p}_3) ]\nonumber
\\
&&+\frac{1}{2}
[ N_{12}B_{ij}(\underline{p}_1)B_{lm}(\underline{p}_2)
-N_{k3}B_{ij}(\underline{p}_3)B_{lm}(\underline{k})
\nonumber
\\
&&
+\tilde N_{31}B_{ij}(\underline{p}_1)B_{lm}(\underline{p}_3)
+\tilde N_{k1}B_{ij}(\underline{p}_1)B_{lm}(\underline{k})
 \label{Cint}
\\
&&
+
\tilde N_{32}B_{ij}(\underline{p}_3) B_{lm}(\underline{p}_2)
+\tilde N_{k2}B_{ij}(\underline{k})B_{lm}(\underline{p}_2) ] \}\nonumber
\eea
where we used $N_{ij}=N_i+N_j$, $\tilde N_{ij}=N_i-N_j$ $(i,j=1,2,3,k)$. 
This  equation can be solved self consistently for the quantity
$C_{ijlm}(\underline{k})$ using the result for $B_{ij}(\underline{k})$ from
(\ref{Bint}).

%\section{Viscosity from Field Theory}
%  Throughout this
%section we use capital letters to denote four-vectors and small letters for
%three-vectors. We also define $\int d^4p/(2\pi)^4:= \int dP $.

Now we turn to response calculation \cite{Hosoya}. We work with the
density matrix in the Heisenberg representation which satisfies,
\be
\frac{\partial \rho}{\partial t}=0
\ee
and  write $\rho=e^{-A+B}/ {\rm Tr} e^{-A+B}$
where $A =\int d^3 x F^\nu T_{0\nu}$ and
$B(t)=\int d^3 x \int _{-\infty}^t dt' e^{\epsilon (t'-t)}T_{\mu\nu}(x,t')
\partial^\mu F^\nu(x,t')$
with $F^\mu=\beta u^\mu$ and $\epsilon$ to be taken to zero at the end. Here
 $A$ is the equilibrium part of the Hamiltonian and $B$ is a
perturbative contribution. We expand  the density matrix:
\bea
\rho&&=\rho_0 \left[ \right. 1+\int_0^1d\lambda( B(\lambda) -\langle
B(\lambda)\rangle )
\nonumber
\\
&&
 +\int_0^1 d\lambda \int_0^\lambda d\tau (B(\lambda)
B(\tau)-\langle B(\lambda) B(\tau)\rangle  )-
\label{exrho}
\\
&&\int_0^1 d\lambda \int_0^1 d\lambda'(\langle  B(\lambda)\rangle
B(\lambda')-\langle  B(\lambda)\rangle \langle  B(\lambda')\rangle
\left.\right] + {\cal O}(B^3) \nonumber
\eea
where $\rho_0=e^{-A}/{\rm Tr} e^{-A}$ is the local equilibrium density matrix.

%\subsection{Viscosity as an Expansion in Green Functions of Composite
%Operators}

%\subsubsection{Linear Response}

Using the first three terms of (\ref{exrho}) produces the linear response 
approximation 
\cite{mhrkb},
\bea
\delta \langle  \pi_{\mu\nu}\rangle^l  = \frac{H_{\mu\nu}}{10}\int d^3 x'
\int_{-\infty}^{t}dt'e^{\epsilon (t'-t)} \int_{-\infty}^{t'}d
t''D_R(x,t;x',t'')
\nonumber
\eea
where $D_R(x,t;x',t'')=-i\theta(t-t'')\langle [\pi(x,t),\pi(x',t'')]\rangle$.
Extracting the shear viscosity using the definition (\ref{DEF})
we obtain in momentum space
\bea
\eta^{(1)} = \frac{1}{10}\frac{d}{d q_0}{\rm Im} [ \lim_{\vec{q} \to
0}D_R(Q)]|_{q_0=0}\,. \label{pl}
\eea
This is the well known Kubo formula \cite{kubo,Hosoya}.

Now we consider corrections to the linear response approximation
\cite{jackiw,blz}.  We calculate
the quadratic shear viscous response from the terms in (\ref{exrho}) that are
quadratic in the interaction. After a lengthy calculation we find that the 
result can be written as a
retarded three-point correlator \cite{mhrkb}:
\bea
&&\delta \langle  \pi_{\mu\nu}(x,t)\rangle ^{q}
=\eta^{(2)} H_{\mu\nu}^{T2} \label{pq} \\
&&\eta^{(2)} = \frac{3}{70}\frac{d}{d q_0} \frac{d}{ dq_0'}{\rm Re}\,[
\lim_{\vec{q} \to 0}G_{R1}(-Q-Q',Q,Q')]|_{q_0=q'_0=0} \nonumber
\eea
with  $G_{R1}(x,y,z)
= \theta(t_x-t_y)\theta(t_y-t_z)\langle[[\pi(x),\pi(y)],\pi(z)]\rangle $ $ +
\theta(t_x-t_z)\theta(t_z-t_y)\langle [[\pi(x),\pi(z)],\pi(y)]\rangle$.  
This is an inter.  We have obtained a type of nonlinear Kubo
formula that allows us to obtain the quadratic shear viscous response from a
retarded three-point function using equilibrium quantum field theory.

Next
we obtain a perturbative expansion for the correlation functions of
composite operators
 $D_R(x,y)$ and $G_{R1}(x,y,z)$ which appear in (\ref{pl}) and (\ref{pq}).
We use the CTP formulation of finite
temperature field theory, and work in the Keldysh representation
\cite{keld,Chou}.  We
define the vertices
$\Gamma_{ij}$ and $M_{ijlm}$ by
truncating external legs from the following connected vertices:
\bea
&& \Gamma^C_{ij} = \langle T_c \pi_{ij}(x) \phi(y) \phi(z)\rangle  \nonumber \\
&& M^C_{ijlm} = \langle T_c \pi_{ij}(x) \pi_{lm}(y)\phi(z) \phi(w)\rangle
\eea
where $\pi_{ij}(x) = \partial_i\phi(x) \partial_j\phi(x) - \frac{1}{3} \delta_{ij}(\partial_m \phi(x))(\partial_m \phi(x))$,
and $T_c$ is  the time ordering operator on the CTP contour.
These definitions allow us to write the two- and three-point correlation
functions as integrals of the form depicted in Fig. ~[1].
\vspace*{-1.5cm}
%%%%%%%%%%%%%%%%%%%%%%%%%%%%% Fig. 1 %%%%%%%%%%%%%%%%%%%%%%%%%%%%%%%%%%%%%
\begin{eqnarray}
\parbox{7cm}
{{
\begin{center}
\parbox{5cm}
{
\epsfxsize=4cm
\epsfysize=2.5cm
\epsfbox{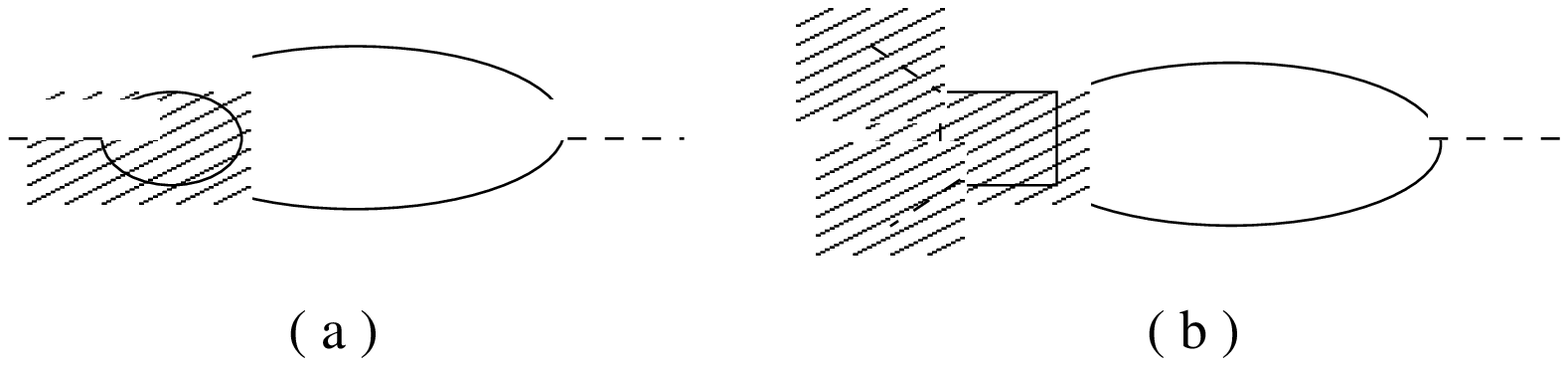}}\\
\parbox{7cm}{\small   Fig.~[1]: (a) Two-point function
 for shear viscosity from linear response; (b) Three-point function for
quadratic shear viscous response. The dashed external line represents the
composite operator $\pi_{ij}$. The square box is the four-point function 
$M$and the round blob is the three-point vertex $\Gamma$.  }
%%%%%%%%%%%%%%%%%%                           %%%%%%%%%%%%%%%%%%%%%%%%%%
%%%%%%%%%%%%%%%%%%%%%%%%%%%%%%%%%%%%%%%%%%%%%%%%%%%%%%%%%%%%%%%%%%%%%%%
%\label{F2}
\end{center}
}}
\nonumber
\end{eqnarray}

After performing
the sum over Keldysh indices using the Mathematica program described in
\cite{johnS} we obtain,
\bea
&& D_R(Q) = i\int dK (N_{k+q} - N_k) \Gamma_{R2}^{ij}(K,Q,-K-Q)
\nonumber
\\
&&~~ D_A(K) D_R(K+Q)
I_{ji}(k) \label{kubo1} \\
&& G_{R1}(-Q-Q',Q,Q') 
=  -4\int dK (\bar M_F)_{ikkj}(K,Q,Q')
\nonumber
\\
&&~~ D_R(K) D_A(K+Q+Q')
I_{ji}(k)  \label{kubo2}
\eea
where  $\bar  M_F = M_F + N_1 M_{R4} + N_4 M_{R1}$ 
is a particular combination of four-point vertices  \cite{johnS2}.
 Rotational invariance leads to
$M_{ijlm} := \hat I_{ij} \hat I_{lm} M$;  $\Gamma_{ij} := \hat I_{ij} \Gamma$.
  We regulate the pinching singularity with the imaginary part of the HTL 
self energy $\Sigma_k$ and obtain \cite{meg},
\bea
D_R(K)D_A(K+Q) \rightarrow  -\frac{\rho_k}{2\rm{Im} \Sigma_k}\label{pinch}
\eea
with $\rho_k = i(D_R(K) - D_A(K))$.
Now we expand in $q_0$ and $q'_0$. In (\ref{kubo1}) and (\ref{kubo2})
 we keep terms proportional to $q_0$ and $q_0 q'_0$ respectively,
 since these terms are the only ones that contribute to (\ref{pl}) and 
(\ref{pq}). Substituting (\ref{kubo1}) and (\ref{kubo2}) into (\ref{pl}) 
and (\ref{pq})  we obtain,
\bea
&& \eta^{(1)} = \frac{\beta}{15} \int dK\, k^2 \,\rho_k n_k (1+n_k) \left[\frac
{\rm Re \Gamma_{R2}(K)}{\rm Im \Sigma_k}\right] \label{ftf11} \\
&& \eta^{(2)} =  -\frac{2\beta^2}{105} \int dK \,k^2\rho_k n_k (1+n_k) N_k
\left[ \frac{\rm Re M_{R1}(K)}{\rm Im \Sigma_k} \right]\,. \label{ftf}
\eea
Comparing with (\ref{ttf11}) and (\ref{ttf}) we see that the results are
identical if we identify
\bea
B(\underline{k}) = \frac {\rm Re \Gamma_{R2}
(\underline{k})}{\rm Im \Sigma_k},\, ~~~
 C(\underline{k}) = -  \frac{\rm Re M_{R1}(\underline{k})}{\rm Im \Sigma_k}
\label{EQ2}
\eea
with the momentum $K$ on the shifted mass shell: $\delta(K^2-m_{th}^2)$ where
$m_{th}^2=m^2+{\rm Re} \Sigma_K$.

It is well known that ladder diagrams  give the
dominate contributions to the vertex $\Gamma_{ij}$. They
 contribute to the viscosity to the same order in perturbation
theory as the
bare one loop graph and thus need to be included in a resummation.
The integral equation that one obtains from resumming ladder contributions 
to the three-point vertex (see Fig.[2])
has exactly the same form as the equation obtained from the
linearized Boltzmann equation (\ref{Bint}) with a shifted mass shell 
describing effective thermal excitations \cite{jeon1,meg}.
\vspace*{-1.7cm}
%%%%%%%%%%%%%%%%%%%%%%%%%%%%% Fig. 2 %%%%%%%%%%%%%%%%%%%%%%%%%%%%%%%%%%%%%
\begin{eqnarray}
\parbox{7cm}
{{
\begin{center}
\parbox{5cm}
{
\epsfxsize=4cm
\epsfysize=2.5cm
\epsfbox{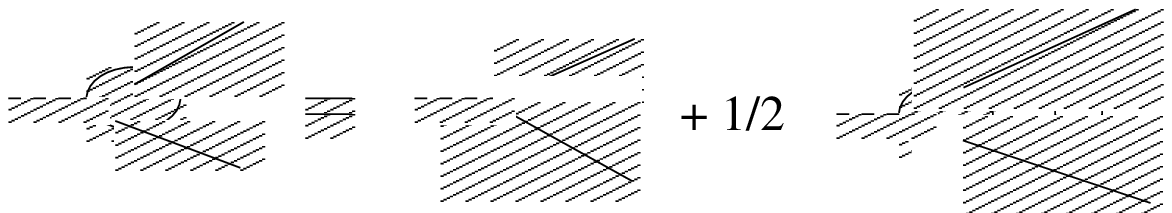}}\\
\parbox{7cm}{\small   Fig.~[2]: Integral equation for the ladder resummation. }
%%%%%%%%%%%%%%%%%%                           %%%%%%%%%%%%%%%%%%%%%%%%%%
%%%%%%%%%%%%%%%%%%%%%%%%%%%%%%%%%%%%%%%%%%%%%%%%%%%%%%%%%%%%%%%%%%%%%%%
%\label{F2}
\end{center}
}}
\nonumber
\end{eqnarray}

Following the pinch effect argument\cite{jeon1,mhrkb}, one can show that an
infinite set of ladder graphs and some other 
contributions which we will call extended ladder graph contribute
to the same order to  vertex   $M_{ijlm}$ as the tree diagram. Therefore,
for consistent calculation, we consider an integral equation which resums
all of these diagrams,  as shown in Fig. ~[3].
%%%%%%%%%%%%%%%%%%%%%%%%%%%%%%%%%%%%%%%%%%%%%%%%%%%%%%%%%%%%%%%%%%%%%%%%%%%%
%%%%%%%%%%%%%%%%%%%% FIG 3 %%%%%%%%%%%%%%%%%%%%%%%%%%%%%%%%%%%%%%%%%%%%%%%%
\vspace*{-1.7cm}
\begin{eqnarray}
\parbox{7cm}
{{
\begin{center}
\parbox{5cm}
{
\epsfxsize=4cm
\epsfysize=3cm
\epsfbox{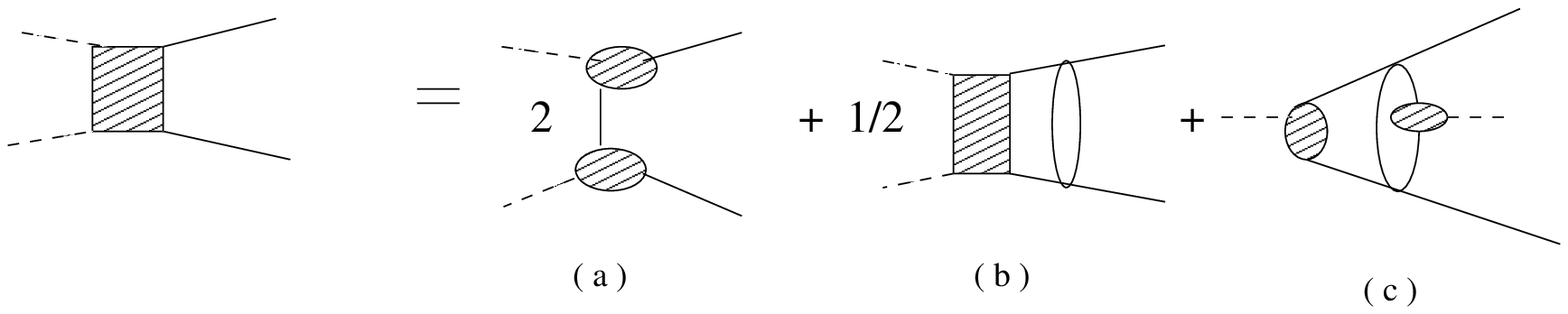}}\\
\parbox{7cm}{\small  Fig.~[3]:  Integral equation for an
extended-ladder resummation.  }
\label{F4}
\end{center}
}}
\nonumber
\end{eqnarray}
We keep only
the pinching contributions, using (\ref{pinch}) to regulate, and  expand in 
$q_0$ and $q_0'$ keeping only  the term proportional to
$q_0q_0'$, since that is the only term that will contribute to the quadratic
shear viscous response coefficient. We obtain \cite{mhrkb}:
\bea
&& N_k  M_{R1}^{ijlm}(K) = - N_kI_{ij} \frac {\Gamma_{R2}^{lm}(K)}{{\rm Im}
\Sigma_k} \\ 
&&~~~~+\frac{\lambda^2}{4}
\int \,dP\,dR\,d_n\rho_p\rho_p'\rho_r
 \bigl[
- \frac {N_{p} M_{R1}^{ijlm}(P)}{{\rm Im}\Sigma_{p}}
\nonumber \\
&&~~~~
+\frac{1}{2} \frac {\Gamma_{R2}^{ij}(P)}{{\rm Im}\Sigma_{p}}\frac
{\Gamma_{R2}^{lm}(R)} {{\rm Im}\Sigma_{r}} \tilde N_{rp}
 + \frac{1}{2}\frac {\Gamma_{R2}^{ij}(P)}{{\rm Im}\Sigma_{p}}\frac
{\Gamma_{R2}^{lm}(K)} {{\rm Im}\Sigma_k} \tilde N_{pk}
\bigr] \,. \nonumber
\eea

We introduce the symmetric notation: $P:=P_1\,$;$~P':=P_2\,$;$~R:=P_3$ and 
rewrite
the equation above after symmetrizing on the integration variables.
We obtain,
\bea
&& N_kI_{ij} \frac{\Gamma_{R2}^{lm}(K)}{{\rm Im}\Sigma_k} 
=\frac{\lambda^2}{12} \int dP_1dP_2dP_3{(2\pi)}^4\delta^4_{(P_1+P_2-P_3-K)}
\nonumber
\\
&&~~~
d_n \rho_1 \rho_2 \rho_3
[\frac{N_{p_3} M_{R1}^{ijlm} (P_3) }{{\rm Im}\Sigma_{p_3}}
+\frac{N_{k} M_{R1}^{ijlm} (K) }{{\rm Im}\Sigma_k}
\nonumber
\\
&&~~~
 - \frac{N_{p_1} M_{R1}^{ijlm} (P_1) }{{\rm Im}\Sigma_{p_1}}
- \frac{N_{p_2} M_{R1}^{ijlm} (P_2) }{{\rm Im}\Sigma_{p_2}}
\label{ieq2}
\\
&&~~~
+\frac{1}{2}\{ N_{12}\frac{\Gamma_{R2}^{ij}(P_1)}{{\rm
Im}\Sigma_{p_1}}
\frac{\Gamma_{R2}^{lm}(P_2)}{{\rm Im}\Sigma_{p_2}}
-N_{k3}\frac{\Gamma_{R2}^{ij}(K)}{{\rm Im}\Sigma_k}
\frac{\Gamma_{R2}^{lm}(P_3)}{{\rm Im}\Sigma_{p_3}}
\nonumber
\\
&&~~~~+\tilde N_{31}\frac{\Gamma_{R2}^{ij}(P_1)}{{\rm Im}\Sigma_{p_1}}
\frac{\Gamma_{R2}^{lm}(P_3)}{{\rm Im}\Sigma_{p_3}}
+ \tilde N_{k1}\frac{\Gamma_{R2}^{ij}(P_1)}{{\rm Im}\Sigma_{p_1}}
\frac{\Gamma_{R2}^{lm}(K)}{{\rm Im}\Sigma_k}
\nonumber
\\
&&~~~~+ \tilde N_{32}\frac{\Gamma_{R2}^{ij}(P_3)}{{\rm Im}\Sigma_{p_3}}
\frac{\Gamma_{R2}^{lm}(P_2)}{{\rm Im}\Sigma_{p_2}}
+ \tilde N_{k2}\frac{\Gamma_{R2}^{ij}(K)}{{\rm Im}\Sigma_k}
\frac{\Gamma_{R2}^{lm}(P_2)}{{\rm Im}\Sigma_{p_2}}
    \}]\,.
\nonumber
\eea
Note that once again we have obtained an integral equation that is decoupled:
it only involves $M_{R1}$ and $\Gamma_{R2}$.  With $\Gamma_{R2}$
determined from the integral equation for the ladder resummation, (\ref{ieq2})
can be solved to obtain $M_{R1}$.
  Finally, comparing (\ref{ftf}) and (\ref{ieq2}) with (\ref{ttf}) and
(\ref{Cint}) we see that calculating the quadratic shear viscous response using
 transport theory describing effective thermal excitations and keeping terms
that are quadratic in the gradient of the four-velocity field in the expansion
of the Boltzmann equation is equivalent to calculating the same response
coefficient from quantum field theory at finite temperature using the
next-to-linear response Kubo formula with a vertex given by a specific integral
equation. This integral equation shows that the complete set of diagrams that
need to be resummed includes the standard ladder graphs, and an additional set
of extended ladder graphs. Some of the diagrams that contribute to the
viscosity are shown in Fig.~[4].
\vspace*{-1.5cm}
\begin{eqnarray}
\parbox{7cm}
{{
\begin{center}
\parbox{5cm}
{
\epsfxsize=4cm
\epsfysize=3cm
\epsfbox{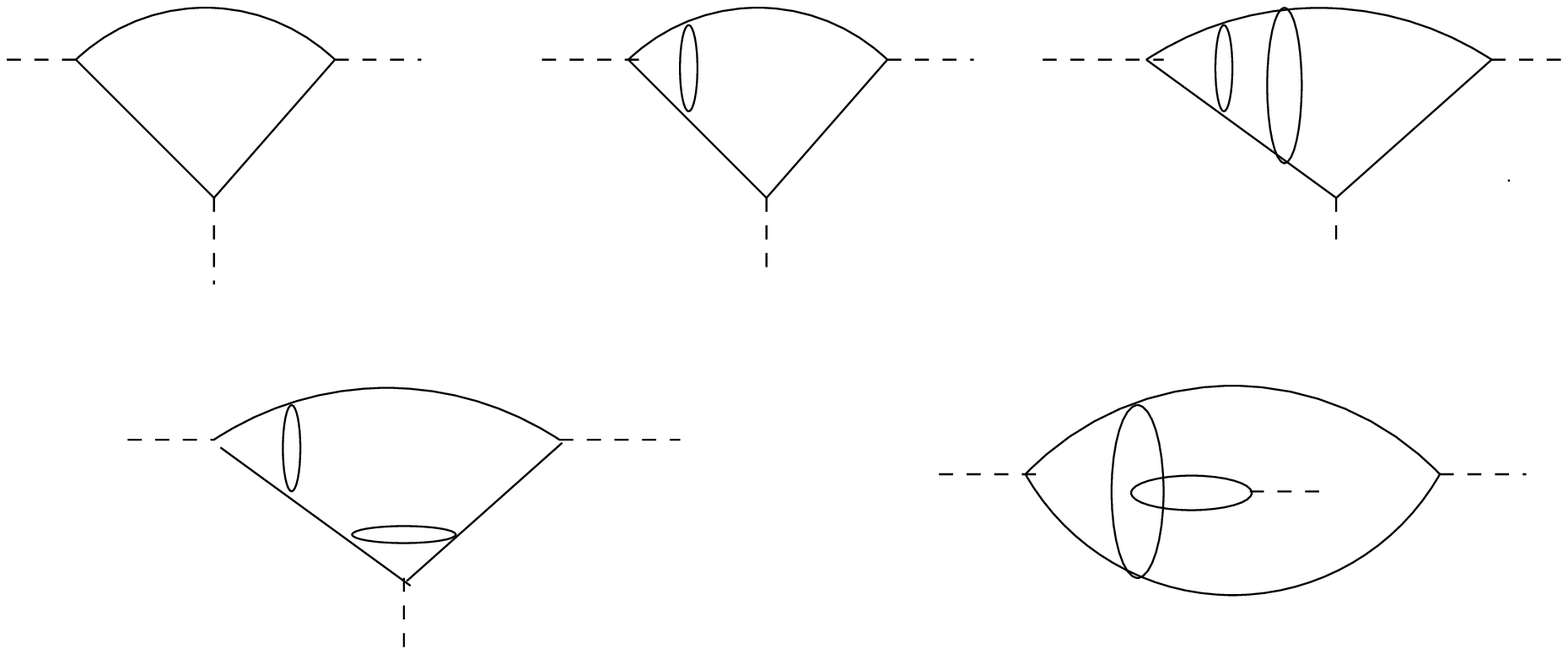}}\\
\parbox{7cm}{\small  Fig.~[4]:  Some of the ladder and extended ladder
diagrams that contribute to quadratic shear viscous response. }
\label{F7}
\end{center}
}}
\nonumber
\end{eqnarray}
This result provides a diagrammatic interpretation of the Chapman-Enskog 
expansion of Boltzmann equation, up to quadratic order.

There are several directions for future work.  It has  been shown that the
Boltzmann equation can be derived from the Kadanoff-Baym equations by using a
gradient expansion and keeping only linear terms \cite{SL}.  The connection
between this result and the work discussed in this paper can probably be
understood by studying the dual roles of the gradient expansion and the
quasiparticle approximation.  In addition, it would be interesting to
generalize this work to gauge field theories.


\vspace*{1cm}

\begin{thebibliography}{}
\bibitem{Groot}
        S.R.~de Groot, W.A.~van Leeuwen, and Ch.G.~van Weert,
        {\it Relativistic Kinetic Theory}, (North-Holland Publishing, 1980).
%\bibitem{hz} U. Heinz, Phys. Rev. Lett. {\bf 51},351 (1983).
%\bibitem {bg}
%V.~A.~Rubakov and M.~E.~Shaposhnikov, Usp.\ Fiz.\ Nauk {\bf 166}, 493 (1996)
%[hep-ph/ 9603208].

\bibitem {hvi} See, for example,
D.~Teaney and E.~V.~Shuryak,  Phys.\ Rev.\ Lett.\  {\bf 83}, 4951 (1999).



\bibitem{MartinK}G. Baym, et al  , Phys.\ Rev.\ Lett.\  {\bf 64}, 1867 (1990).

\bibitem{kubo}
   D.N. Zubarev, {\it Nonequilibrium Statistical Thermodynamics},
   (Plenum, New York, 1974).

 \bibitem{Hosoya} A.~Hosoya, M.~Sakagami, and M.~Takao,
                   Ann.~of Phys.   (NY) {\bf 154}, 229 (1984),
                  and references therein.

\bibitem{jeon1} S. Jeon, Phys.
Rev. {\bf D52}, 3591 (1995); S.~Jeon and L. Yaffe,
                Phys.~Rev.~D {\bf 53}, 5799 (1996).



  \bibitem{pisarski}
   R.D. Pisarski, Phys. Rev. Lett. {\bf 63}, 1129 (1989); E. Braaten
   and R.D. Pisarski, Nucl. Phys. B {\bf 337}, 569 (1990).

\bibitem{LAD} V.V. Lebedev and A.V. Smilga, Physica A {\bf 181}, 187 (1992);
M.E. Carrington, Phys. Rev. {\bf D48}, 3836 (1993).

\bibitem {bdc}
D.~Bodeker, Phys.\ Lett.\  {\bf B426}, 351 (1998); 
D. F. Litim and C. Manuel, Phys. Rev. Lett. {\bf 82}, 4981 (1999).
\bibitem{jackiw} 
 R. Jackiw and V.P. Nair,Phys. Rev. {bf D48}, 4991 (1993).
\bibitem{blz} J -P Blaizot and E. Iancu,   hep-ph/0101103 .
% Nucl. Phys. B {\bf 570}, 326 (2000);
\bibitem {ASY} P.~Arnold, D.~T.~Son and L.~G.~Yaffe,
Phys.\ Rev.\  {\bf D59}, 105020 (1999).

\bibitem{mhrkb} M.E. Carrington, Hou Defu and R. Kobes, hep/ph 0102256.

\bibitem{meg} M.E. Carrington, Hou Defu and R. Kobes, 
Phys. Rev. {\bf D62}, 025010 (2000).


\bibitem{keld} L.V. Keldysh, Zh. Eksp. Teor. Fiz. {\bf 47}, 1515 (1964).

\bibitem{Chou}
  K.-C. Chou, Z.-B. Su, B.-L. Hao, and L. Yu, Phys. Rep. {\bf 118},
  1 (1985).


\bibitem{johnS}   M.E. Carrington, Hou Defu, A. Hachkowski, D. Pickering and J.
C. Sowiak,  Phys. Rev.  {\bf D61}, 25011 (2000).

\bibitem{johnS2} M.E. Carrington, Hou Defu and J.C. Sowiak, 
Phys. Rev. {\bf D62}, 065003, (2000).

\bibitem{SL} S. Leupold, Nucl. Phys. A {\bf 672}, 475, (2000).


\end{thebibliography}
\end{document}